\documentclass[12pt]{article}

\usepackage{epsfig}

\usepackage{amssymb}

\usepackage{amsmath}
\usepackage{amsfonts}

\usepackage{amsthm}

\usepackage{color}

\theoremstyle{theorem}

\theoremstyle{definition}

\def\bp{\begin{proof}}
\def\ep{\end{proof}}

\def\be{\begin{equation}}
\def\ee{\end{equation}}
\def\ba{\begin{array}{c}}
\def\ea{\end{array}}

\def\ben{$$}
\def\een{$$}

\newcommand{\bea}{\begin{eqnarray}}
\newcommand{\eea}{\end{eqnarray}}

\begin{document}

\titlepage


 \begin{center}{\Large \bf


Quantum star-graph analogues of ${\cal PT}-$symmetric square wells.
II: Spectra.

  }\end{center}

\vspace{4mm}

 \begin{center}

 {\bf Miloslav Znojil}

 \vspace{2mm}
Nuclear Physics Institute ASCR,

250 68 \v{R}e\v{z}, Czech Republic

{e-mail: znojil@ujf.cas.cz}

\vspace{2mm}

\end{center}

\vspace{5mm}

\section*{Abstract}

For non-Hermitian equilateral $q-$pointed star-shaped quantum graphs
of paper I [Can. J. Phys. 90, 1287 (2012), arXiv 1205.5211] we show
that due to certain dynamical aspects of the model as controlled by
the external, rotation-symmetric complex Robin boundary conditions,
the spectrum is obtainable in a closed asymptotic-expansion form, in
principle at least. Explicit formulae up to the second order are
derived for illustration, and a few comments on their consequences
are added.



%

\vspace{5mm}

{\large \bf PACS}

03.65.Ca (Formalism), 03.65.Db (Functional analytical methods),

03.65.Ta (Foundations of quantum mechanics; measurement theory) and

03.70.+k (Theory of quantized fields)

\newpage


\section{Introduction}

\begin{figure}[h]                     
\begin{center}                         
\epsfig{file=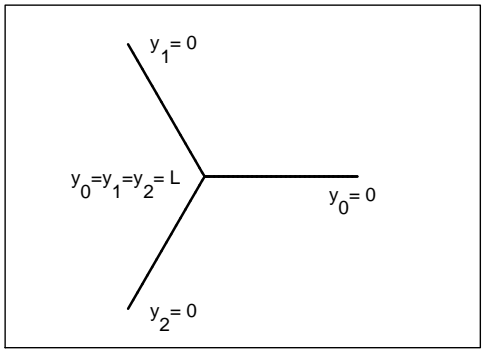,angle=270,width=0.4\textwidth}
\end{center}                         
\vspace{-2mm} \caption{Star-shaped-graph coordinates running inwards
($q=3$).
 \label{xjedu}}
\end{figure}

In paper I \cite{canad} we recalled the recent growth of popularity
of the study of quantum systems in their ${\cal PT}-$symmetric {\it
alias} pseudo-Hermitian representations (cf., e.g., review papers
\cite{Carl} and \cite{ali}, respectively). In this broader context
we summarized briefly the basic characteristics of one of the
exactly solvable illustrative examples of such a quantum system. We
emphasized that the model proposed in \cite{sdavidem} proved useful,
as a mathematically consistent methodical laboratory, in a few
subsequent publications (cf., e.g., their summaries in
\cite{davidia}).
Next, we reinterpreted the latter toy model (which lives on a finite
real interval $(-L,L)$) as one of the most elementary (viz.,
two-pointed-star-shaped) exemplifications of a non-Hermitian quantum
graph \cite{[6a]}. We felt inspired by the friendly nature and tractability of
the model and we proposed one of its most natural  $q-$pointed
star-shaped generalizations with $q>2$  (cf. the first nontrivial
sample graph in Fig.~\ref{xjedu}).

%

The main result of paper I was the derivation of a universal and
compact form of the secular equation which defined the spectra in
implicit form. We also added a few mathematics- as well as
physics-oriented overall comments on the spectrum and, in
particular, we employed a brute-force, purely numerical method in
order to demonstrate that at $q>2$, some of the energies may prove
complex in general.

In the present brief completion of paper I we intend to pay more
attention to the symmetries of the model. In essence, we shall be
able to reveal that the formal and algebraic boundary-condition
symmetries as carried by the Hamiltonian really exhibit a nontrivial
connection with certain approximate symmetries of the geometry of
certain subsets of the energy roots in complex plane.

\section{The model}


Let us accept the convention that the family of the star-shaped
graphs $G^{(q)}$ is numbered by an integer $q=\,(2,)\,3,4,\ldots\ $.
and that in each member of this family the $j-$th edge is
parametrized by the respective coordinate $y_j$. Let us also set
this coordinate equal to zero at the external end of the edge, and
let it grow up to value $y_j=L$ at the central vertex (cf.
Fig.~\ref{xjedu} where $q=3$).

For the whole family of the equilateral star-shaped non-Hermitian
quantum graphs $\mathbb{G}^{(q)}$ we postulated, in paper I, {a
Laplace-type operator on the graphs in the form which generalizes
the Hamiltonian of Ref.~\cite{sdavidem} and which was introduced
via} a $q-$plet of its ordinary differential free-motion components,
 \be
 -\frac{d^2}{dy_{j}^2}\psi_{j}(y_{j})
 = k^2\,\psi_{j}(y_{j})
 \,,\ \ \ \ \ \ \
 y_{j}\in (0,L)\,,
 \ \ \ \ \ j = 0, 1, \ldots, q-1\,.
 \label{kveplich}
 \ee
The Kirchhoff's continuity law was imposed to match the flows and
wave functions at the central vertex,
 \begin{equation}
 \label{bcinge}
    \sum_{m=0}^{q-1}\,\partial_{y_{m}}\psi_{m}(L)=0\,,
    \ \ \ \ \
 \psi_{j}(L)=\psi_{0}(L)\,,\ \ \ \ j = 1,2, \ldots, q-1
  \,.
 \end{equation}
At the outer ends of the edges we picked up the manifestly
non-Hermitian boundary conditions
 \begin{equation}
 \label{brody}
  \partial_{x_{j}}\psi_{j}(0)= i \alpha\, e^{{\rm i}j\varphi}\,
  \psi_{j}(0)\,,
  \ \ \ \ \ j = 0, 1, \ldots, q-1\,,
  \ \ \ \varphi=2\pi/q
  \,,\ \ \ \ \ \alpha>0
 \end{equation}
exhibiting, at any fixed $q$, {a constant growth of phase with a
move $j \to j+1$ to} the neighboring edge.

\subsection{Secular equation}

Due to {the simplicity of} Eq.~(\ref{brody}) we were able to prove,
in paper I, the existence of infinitely many bound states of the
system $\mathbb{G}^{(q)}$ with real energies. In implicit form,
these energies were defined via the real roots $\kappa=k\,L$ of
secular equation
 \be
  \label{bqs}
  \frac{
 \kappa^{p+2}+({\rm i}\beta)^{p+2}\,
 \tan^{p} \kappa}{\kappa^{p+2}-({\rm i} \beta)^{p+2} \tan^{p+2} \kappa
    }\,\tan \kappa
    =0\,,
    \ \ \ \ \
   {\beta=\alpha\,L\,,}
  \ \ \ \
    p = q-2 = 0, 1, \ldots\,.
 \ee
For non-trivial quantum stars with positive $p$ the implicit energy
formula (\ref{bqs}) was the main result of  paper I. We also showed
there that although the two free parameters $L$ and $\alpha$  of the
model were chosen real and positive, some of the energy roots might
still be admitted complex in general, $\kappa \in \mathbb{C}$.

In paper I there was no space left for a more detailed
analysis of the $p>0$ spectra. In what follows we intend to fill the
gap and to turn attention to the open questions of the existence,
type and parametric dependence of {\em all } of the roots
$\kappa_n(\beta)\,$ of Eq.~(\ref{bqs}), real or complex.

\subsection{Two sets of the roots of
secular equation}

A subset of the roots of Eq.~(\ref{bqs}) will originate from the
zeros of function $\tan \kappa$ {(via the second factor of the
left-hand-side expression)} and from the poles of function $\tan
\kappa$ {[due to the higher power (viz., $p+2$) of this function in
the denominator of the first factor of the left-hand-side
expression]}. This subset is real, $\beta-$independent, equidistant
and may be $^{[1]}-$superscripted,
 \be
  \kappa=\kappa_n^{[1]}=n\pi/2\,,\ \ \ \ \ n=1,2,\ldots\,.
  \label{realp}
 \ee
The remaining, $\beta-$dependent and, in general, complex roots $
\kappa=\kappa^{[2]}(\beta)=\lambda$ should be then sought via the
remaining, reduced part of the secular equation,
 \be
  \label{uqs2}
 \tan^{p} \lambda
   =-\left (\frac{ \lambda}{{\rm i}\beta}\right )^{p+2}\,.
 \ee
At any positive $p=1,2,\ldots$
the latter transcendental
relation is equivalent to the $p-$plet of
simpler equations
 \be
  \label{us23}
 \tan  \lambda_{[n]}
    =\left (\frac{ \lambda_{[n]}}{\beta}\right )^{1+2/p}
    \,{\rm e}^{ {\rm i}\pi\left (-\frac{1}{2}+\frac{2n}{p}\right )}\,,
    \ \ \ \ \ \ \ n = 0, 1, \ldots, p-1\,.
 \ee
In spite of the scepticism as expressed in  paper I (cf.
also a few complementary comments on the localization of non-real
roots in proceedings \cite{actacan}), this equation is still
amenable to non-numerical analysis. This will be our present key
observation.

\section{The second, complex subset of the roots}

\subsection{Non-numerical localization}

At a fixed $\beta>0$ and at the larger values of the imaginary part
of $ \lambda_{[n]}$ the absolute value of the left-hand-side
function lies very close to one so that
the absolute value of $ \lambda_{[n]}$ cannot in fact be too large.
{\it Vice versa}, at the larger absolute values of $ \lambda_{[n]}$
the roots of Eq.~(\ref{us23}) cannot lie too far from the real axis
and, in particular, from the poles of the left-hand-side function.
Thus, it makes sense to reparametrize
 \be
 \lambda_{[n]}=\kappa^{[2]}_{[n]}(\beta)
 =\frac{2M+1}{2}\,\pi+\varepsilon_{[n]}(\beta,M)\,,\ \ \ \ M = 0, 1,
 \ldots\,.
  \label{circularp}
 \ee
Whenever the auxiliary integer $M$ is chosen sufficiently large,
quantities $\varepsilon_{[n]}(\beta,M)$ may be assumed small. Such
an assumption is not inconsistent since after the $M-$dependent
change  $\lambda \to \varepsilon$ of our variables one obtains
another form of secular Eqs.~(\ref{us23}),
 \be
  \label{us235}
 {\tan \varepsilon_{[n]}}
    =\left (\frac{\beta}{(M+1/2)\pi+\varepsilon_{[n]}}
    \right )^{1+2/p}
    \,{\rm e}^{- {\rm i}\pi\left (\frac{1}{2}+\frac{2n}{p}\right
    )}\,,
    \ \ \ \ \ \ \ n = 0, 1, \ldots, p-1\,.
 \ee
From this exact implicit relation we may derive explicit approximate
formula
 \be
  \label{us2358}
 { \varepsilon_{[n]}}
    =\left (\frac{\beta}{(M+1/2)\pi}\right )^{1+2/p}
    \,{\rm e}^{- {\rm i}\pi\left (\frac{1}{2}+\frac{2n}{p}\right )}
    \left [1+{\cal O}\left(\frac{1}{M^{1+2/p}}\right)\right ]\,
 \ee
or, on the next level of precision and with an amended error factor,
 \be
  \label{us23589}
 { \varepsilon_{[n]}}
    =\frac{(M+1/2)\pi}{{1+2/p}+\beta^{-1-2/p}[(M+1/2)\pi]^{2+2/p}
  \,{\rm e}^{ {\rm i}\pi\left (\frac{1}{2}+\frac{2n}{p}\right )}{}}\,
    \left [1+{\cal O}\left(\frac{1}{M^{3+4/p}}\right)\right ]\,.
        \ee
These estimates imply that in certain complex discs centered at real
axis we could solve Eq.~(\ref{us235}) by iterations.

{It seems worth adding that the asymptotic-estimate nature of
formulae (\ref{us2358}) or (\ref{us23589}) leaves the question of
the exhaustive description of the complex roots in a non-asymptotic
domain (of $M$) open to further analysis. Perhaps, the comparatively
elementary nature of our secular Eq.~(\ref{bqs}) might provoke a
mathematician to provide a fully rigorous proof of the results of
our present, large$-M$ approximative analysis.}

\subsection{Asymptotic series}

A concrete implementation of the iteration recipe could proceed either
numerically or via algebraic, computer-assisted symbolic
manipulations. In both cases, the qualitative messages delivered by
the construction will be very similar, showing that our
non-Hermitian quantum graphs $\mathbb{G}^{(p+2)}$ support the
existence of $p-$plets of large complex eigenvalues with  small
imaginary parts which lie on approximate circles with very small
radii.

A more explicit characteristics of these circles of eigenvalues may
be deduced from our asymptotic formulae (\ref{us2358}) or
(\ref{us23589}) as well as, if necessary, from their higher-order
asymptotic-series descendants. One may immediately see that these
circles are centered at the odd real eigenvalues
$\kappa_{2M+1}^{[1]}$. Far from the origin we could even assign
every such a $p-$plet a shared integer index $M$ and re-subscript
the eigenvalues yielding $\lambda_{[M,m]}$ with $m=0,1,\ldots,p-1$.

With the decrease of $M$ the radii and deformations of the circles
may cease to be small. Their description becomes numerical. The rate
of convergence of their numerical localization may systematically be
amended via higher-order asymptotic formulae for $\varepsilon$.
During the derivation of these formulae one should keep in mind that
in exact formula (\ref{us235}) the game is entered by the value of
$p=q-2$. This leads to the necessity of more careful algebraic
manipulations.

First of all, the left-hand-side Taylor series must be constructed in
the odd powers of $\varepsilon_{[n]}={\cal
O}\left(\frac{1}{M^{1+2/p}}\right)$ yielding the sequence of
corrections ${\cal O}\left(\frac{1}{M^{3+6/p}}\right)$, ${\cal
O}\left(\frac{1}{M^{5+10/p}}\right)$, etc. The resulting expression
must then match the right-hand-side Taylor series in the integer
powers of $\varepsilon_{[n]}/[(M+1/2)\pi]={\cal
O}\left(\frac{1}{M^{2+2/p}}\right)$ so that {another, different}
sequence of corrections emerges, of the orders of magnitude ${\cal
O}\left(\frac{1}{M^{3+4/p}}\right)$, ${\cal
O}\left(\frac{1}{M^{5+6/p}}\right)$, etc.

In principle, the resulting dedicated double-series semi-numerical
recipe is feasible and straightforward and the resulting formulae
may be easily stored in the computer. Its only unpleasant feature is
that the formulae become too long to be displayed in print.
Nevertheless, in Refs.~\cite{canad} and \cite{actacan} we pointed
out that for the practical numerical purposes of localization of
individual roots $\varepsilon_{[n]}$ at the smallest $M$, even the
generic, non-dedicated numerical algorithms are also able to provide
satisfactory results.

\section{Discussion}

\subsection{The generalized Robin boundary conditions}

One of the features which made the exactly solvable $q=2$ model
$\mathbb{G}^{(2)}$ of Ref.~\cite{sdavidem} truly important in the
context of quantum theory was the ${\cal PT}-$invariance of its
Hamiltonian $H$ with parity ${\cal P}$, time reversal ${\cal T}$ and
symmetry relation $H{\cal PT}={\cal PT}H$. In  paper I we
just pointed out that even such an elementary $q=2$ model
illustrates, in parallel, also several important features of a
non-Hermitian quantum graph. Thus, taking the $q=2$ quantum model
$\mathbb{G}^{(2)}$ as a guide to its $q>2$ generalization, we
replaced the original Schr\"{o}dinger equation of
Ref.~\cite{sdavidem} (living on an interval) by its two-subinterval
version, with a single oriented edge $(\to)$ replaced by two edges,
$(\to,\to)$. Besides such an oriented-graph reinterpretation  it was
necessary to add the explicit Kirchhoff's condition of smoothness of
wave functions in the origin. Finally, we came to the $q=2$ graph
$G^{(2)}=(\to,\leftarrow)$ of the present paper via a re-orientation
of one of the edges/subintervals.

The first consequence of such a change of the traditional
presentation of the model was that the action of the parity  ${\cal
P}$ has been simplified, mediating just a mutual interchange of the
two edges of the graph. This operator may be easily generalized to
${\cal P}^{(q)}$ which replaces each edge $e$ by its left neighbor,
i.e., with $e_j \to e_{j+1}$ and $e_{q-1} \to e_{0}$ at the end of
course.

The above-mentioned ${\cal PT}-$symmetry of the $q=2$ model may be
traced back to its specific Robin boundary conditions
 \begin{equation}\label{bcr}
  \psi'_{0,1}(0)= \pm i \alpha \;\! \psi_{0,1}(0)\,, \ \ \ \ q=2
 \end{equation}
in which one may compensate the mutual interchange of the two edges
(i.e., the action of ${\cal P}^{(2)}$) by the action of an operator
of time reversal ${\cal T}={\cal T}^{(2)}$. The latter action may be
realized either as complex conjugation (${\cal T}{\rm i} = -{\rm
i}$) or,  in the spirit of Refs.~\cite{noninvolut}, as the
180-degrees rotation in the complex plane of the parameter, (${\cal
T}{\alpha} = -{\alpha}$). In  paper I the above-defined
graphs $\mathbb{G}^{(q)}$ were proposed precisely as one of the most
natural realizations of the idea of the ${\cal P}^{(q)}{\cal
T}^{(q)}-$invariance in its generalization from $q=2$ to all $q>2$.

In  paper I the latter symmetry idea has not in fact been
mentioned at all. One of the main reasons was that at $q>2$, the
alternative definitions of ${\cal T}^{(q)}$ cease to be equivalent.
In this sense the present results on the spectra may be perceived as
a support of the  most easily graph-adapted
complex-rotation-operator definition. When accepted, operators
${\cal T}^{(q)}$ would just cause a clockwise rotation in the
complex plane of $\alpha$ by the $q-$dependent angle
$\varphi=2\pi/q$.

Under this convention the idea of rotations is shared by both its
kinematical implementations (via ${\cal P}^{(q)}$ which realize the
spatial maps between edges) and its dynamical realization (via
${\cal T}^{(q)}$ which is a complex rotation of coupling constants
$\alpha$). Thus, purely formally, we could still keep speaking about
a generalized ${\cal PT}-$symmetry in principle, although,
incidentally, it appeared to remain spontaneously broken in all of
our present specific toy-model quantum graphs. Moreover, in contrast
to the involutive character of the parity at $q=2$ (meaning that its
square is identity, $\left [{\cal P}^{(2)}\right ]^2=I$ so that we
can treat it as a reflection), we only have the weaker rule $\left
[{\cal P}^{(q)}\right ]^q=I$ in general. This means that the
mathematical importance of parity ${\cal P}$ (playing the role of a
Krein-space metric at $q=2$ \cite{Langer}) and of the ${\cal
PT}-$symmetry of the $q=2$ Hamiltonian $H$ (meaning that $H$ is
assumed self-adjoint in such a Krein space) are both lost at $q>2$.

\subsection{Methodical aspects of the model}

In our present paper we found that one of the most unexpected
features of the whole family of the nontrivial, genuine non-line
graph-shaped choices of paper I is the wealth of the complex energy
roots of the secular equation. This discovery should be interpreted
as a word of warning because for $q>2$ the model should be treated
as a sample of a gain-loss system in optics (where the complex
eigenvalues keep carrying a physical meaning) rather than as a
sample of a quantum system with resonances.

One should notice that the majority of the complex energy roots of
our model lies in a very close vicinity of the real line. In the
language of physics, this feature of the spectrum could be perceived
as a possibility of a comparatively weak violation of the stability
and of the unitarity of the evolution, but a detailed analysis of
these interpretation possibilities lies beyond the scope of the
present paper.

In the conceptual setting, the present discovery of the wealth of
the complex eigenvalues may be read both as disappointing and as
promising. What is disappointing is that one cannot treat our
present family of quantum graphs as a certain discrete guide to a
reduction of technical difficulties during transition from the
popular ordinary differential non-Hermitian Hamiltonians to their
more sophisticated partial differential descendants \cite{guides}.

On positive side, our present non-Hermitian quantum graphs
$\mathbb{G}$ still may be perceived as belonging to the simplest
nontrivial non-numerically tractable quantum models. Their
conceptual transparency makes their study rewarding and productive
since several exceptional exact-solvability features of the $q=2$
special case (and, in particular, the real part of the spectrum)
still survived the generalization to $q>2$.

In a historical perspective one can reveal analogies between the
present and other graph-based discretizations of a more-dimensional
phase space, say, with the central-symmetry-based reduction of the
three-dimensional hydrogen atom description to an ordinary radial
Schr\"{o}dinger equation with a local effective interaction
$V_{(Coul)}(r) \sim 1/r$ which lives on a semi-infinite interval of
the radial coordinate $r \in (0,\infty)$.

It is worth adding that the most elementary hydrogen-atom model did
also serve methodical purposes in ${\cal PT}-$symmetric quantum
mechanics (cf., e.g., \cite{ptcoul}). At the same time, the friendly
tractability of its Hamiltonian $\mathfrak{h}$ is lost in the
majority of the modern applications of quantum theory. Incidentally,
precisely this also motivated the acceptance of non-Hermitian
Hamiltonians. For heavy nuclei, for example, a complicated {\em
non-unitary} isospectral preconditioning of the realistic
Hamiltonian
 $$
 \mathfrak{h} \to H = \Omega^{-1} \mathfrak{h} \Omega \neq
 H^\dagger
 $$
appeared necessary for keeping the diagonalization of non-Hermitian
$H$ feasible. This requirement also led to the theoretical
formulation of the recipe of the Hermitization in quantum mechanics
\cite{Geyer}.

\subsection{Symmetries}

One of the main surprises provided by our present results is that
the expected parallels between the $q>2$ and $q=2$ non-Hermitian
star-shaped quantum graphs $\mathbb{G}^{(q)}$ appeared practically
non-existent. The main difference between the $q>2$ and $q=2$ models
was found in the nontriviality of the split of the spectrum in two
infinitely large groups in the former case. Besides the expected,
infinitely many real, dynamics-independent eigenvalues
$\kappa^{[1]}_m$, $m=1,2,\ldots$ which kept existing in both cases,
the single real eigenvalue which appeared dynamics-dependent (i.e.,
$\alpha-$dependent) at $q=2$ \cite{sdavidem} was replaced, at any
$q>2$, by {\em infinitely many} eigenvalues $\kappa^{[2]}(\beta)$.
which were dynamics-dependent and, up to exceptions, {\em complex}.

Such a result was utterly unexpected. We also found that whenever
the absolute values of the latter quantities  $\kappa^{[2]}(\beta)$
prove sufficiently large, we may visualize them as forming
approximatively circular complex $p-$plets (\ref{circularp}) which
encircle every odd (and sufficiently large) real eigenvalue
$\kappa_{2M+1}^{[1]}$. Hence, asymptotically at least, we may number
these roots $\kappa^{[2]}(\beta)$ by a pair of integers $M \ (\gg
1)$ and $m\ (=0,1,\ldots,p-1)$. At a fixed $M$, differences
$\varepsilon_{[M,m]}(\beta)
=\kappa^{[2]}_{[M,n]}(\beta)-\kappa_{2M+1}^{[1]} $ remain small and,
in leading order, $m-$independent, $\varepsilon_{[M,m]}(\beta)={\cal
O}\left(\frac{1}{M^{1+2/p}}\right)$.

Let us now emphasize that this result leads to an interesting
generalization of the concept of ${\cal PT}-$symmetry as exhibited
by the present quantum graphs $\mathbb{G}^{(q)}$. This
generalization is based on the fact that the action of the
``generalized time reversal'' ${\cal T}^{(q)}$ changes the Robin
boundary conditions, i.e., in effect, it causes a rotation of the
supporting graph $G^{(q)}$. Nevertheless, this action may be also
perceived as compensating the opposite rotation of the supporting
graph $G^{(q)}$ as caused by the ``generalized spatial reversal''
${\cal P}^{(q)}$. Thus, in a purely formal setting we may summarize
that a combined action of product ${\cal P}^{(q)}{\cal T}^{(q)}$ of
our above-defined operators leaves the supporting graph $G^{(q)}$
invariant.

{\em Simultaneously and approximatively}, the action of ${\cal
P}^{(q)}{\cal T}^{(q)}$ (or, better, of ${\cal T}^{(q)}$ -- because
${\cal P}^{(q)}$ itself has no effect) also {\em rotates} the
discrete, $p-$point asymptotic {\em complex circles} of the
second-subset leading-order eigenvalues and leaves them invariant.
This observation follows from formula (\ref{us2358}) which defines
the approximate solution at a fixed $M \gg 1$ and implies that
 \ben
 \fbox{{\rm transformation}$\ \ \
 \alpha \to {\cal T}^{(q)} \alpha  = \alpha \,e^{-2\pi i/q}
 \ \ \ ${\rm implies\ transformation}
 $ \ \ \
 \varepsilon_{[n]} \to \varepsilon_{[n+1]}$}\,.
 \een
In other words, the asymptotically dominant part of the effect of
the rotation ${\cal T}^{(q)}$ will just cause the change of the
subscript (or, if you wish, of the phase) of the individual complex
roots $\varepsilon_{[n]}$.

%


\newpage

\begin{thebibliography}{00}

%
%
%
%
%
%
%





\bibitem{canad}
M. Znojil, Can. J. Phys. 90 (2012) 1287.


\bibitem{Carl}
C. M. Bender, Rep. Prog. Phys. 70 (2007) 947.

\bibitem{ali}
A. Mostafazadeh, Int. J. Geom. Meth. Mod. Phys. 7 (2010) 1191.



\bibitem{sdavidem}
D. Krejcirik, H. Bila and M. Znojil,
J. Phys. A 39 (2006) 10143.


\bibitem{davidia}
D.  Krej\v{c}i\v{r}\'{\i}k,
J. Phys. A: Math. Gen. 41 (2008) 244012;

D. Krejcirik, P. Siegl and J. Zelezny,
Complex Anal. Oper. Theory 8 (2014) 255.









\bibitem{[6a]}
A. Hussein, D. Krejcirik and P. Siegl,
Trans. Amer. Math. Soc., to appear. Preprint on arXiv:1308.4264


\bibitem{actacan}
M. Znojil,
Acta Polytech. 53 (2013) 317.

\bibitem{noninvolut}
L. Solombrino, J. Math. Phys. 43 (2002) 5439;

M. Znojil, Phys. Lett. A 353 (2006) 463;

A. Mostafazadeh,
 J. Phys. A: Math. Theor. 41 (2008) 055304.
%
%


\bibitem{Langer}
H. Langer and Ch. Tretter,
Czech. J. Phys. 54 (2004) 1113.


\bibitem{guides}
D. Krejcirik and P. Siegl,
%
J. Phys. A: Math. Theor. 43 (2010) 485204;

D. Borisov and D. Krejcirik,
Asympt. Anal. 76 (2012) 49.
%
%
%


\bibitem{ptcoul}
G. Levai, P. Siegl and M. Znojil,
J. Phys. A: Math. Theor. 42 (2009) 295201.



\bibitem{Geyer}
F. G. Scholtz, H. B. Geyer and F. J. W. Hahne, Ann. Phys. (NY) 213
 (1992) 74.





\end{thebibliography}
\end{document}